\documentclass[conf]{new-aiaa}
\usepackage[utf8]{inputenc}

\usepackage{graphicx}
\usepackage{amsmath}
\usepackage[version=4]{mhchem}
\usepackage{siunitx}
\usepackage{longtable,tabularx}
\usepackage{physics}
\usepackage{wrapfig}
\usepackage[center]{subfigure}
\usepackage{subfigmat}

\newcommand*{\rttensor}[1]{\bar{\bar{#1}}}

\title{Coanda Flow Characterization on Base Bleed Configurations Using Global Stability Analysis}

\author{Alejandro Martinez-Cava\footnote{Early Stage Researcher, Applied Mathematics Department}, Eusebio Valero\footnote{Professor of Applied Mathematics, Applied Mathematics Department} and Javier De Vicente\footnote{Associate Professor of Applied Mathematics, Applied Mathematics Department}}
\affil{School of Aeronautical and Space Engineering, Universidad Politécnica de Madrid, \\Plaza Cardenal Cisneros 3, E-28040, Madrid, Spain}
\author{Guillermo Paniagua\footnote{Associate Professor of Mechanical Engineering, Zucrow Labs, 500 Allison Road, AIAA Associate Fellow}}
\affil{Purdue University, West Lafayette, IN}

\begin{document}

\maketitle

\begin{abstract}
Base pressure control is often employed on drag reduction design, but the interaction of the jet with the base region flow topology can generate undesired or uncontrolled flow configurations. The ejected flow can drive pressure bifurcations at the trailing edge, that without a correct optimization can affect the aerodynamic performance of the system. The purpose of this work is to fully understand the physical mechanisms related with the use of injected base bleed employing global stability analysis and adjoint methodologies. Moreover, the results of the sensitivity analysis are used to identify the regions more receptive to passive and active flow control methodologies. This work has a particular relevance in turbine research, where coolant flow is ejected at the trailing edge to ensure an adequate thermal protection of the blade and the downstream turbine stages.
\end{abstract}

\section*{Nomenclature}

{\renewcommand\arraystretch{1.0}
\noindent
\begin{longtable*}{@{}l @{\quad=\quad} l@{}}
$\Omega$  & control volume \\
$\Omega_i$  & mesh subdomain \\
$\sigma$  & complex eigenvalue \\
$\sigma_i$  & eigenvalue imaginary part (pulsation) \\
$\sigma_r$  & eigenvalue real part (amplification rate) \\
$\nabla_s \sigma$  & structural sensitivity \\
$\nabla_{\mathbf{\bar{q}}} \sigma$  &  sensitivity to base flow modifications \\
$\nabla_{\mathbf{q_f}} \sigma$  &  sensitivity to steady forcing \\
$\nabla_p \sigma$  &  sensitivity to passive control \\
$\mathbf{A}$  & linearized Jacobian matrix \\
$\mathbf{B}$  & sensitivity matrix \\
c  &  model chord length \\
$C_b$  &  base bleed pressure relation coefficient \\
$\rttensor{F}$  & flux tensor \\
$GCI$  &  Grid Convergence Index \\
$\mathbf{M}$  &  volume matrix \\
$N$  & number of grid points \\
$\mathcal{N}_f$  & number of faces on subdomain \\
$N_v$  & number of fluid variables \\
$\mathbf{n}$  & normal direction to the body surface \\
$\mathbf{\bar{q}}$  & \textit{mean flow} vector field \\
$\mathbf{\bar{q}^+}$  & adjoint operator of the \textit{mean flow} vector \\
$\mathbf{q}$  & conservative variables vector \\
$\mathbf{q_i}$  & discretized vector state solution in subdomain $\Omega_i$ \\
$\mathbf{\widetilde{q}}$  & vector of small perturbations \\
$\mathbf{\widehat{q}}$  & direct global mode vector \\
$\mathbf{\widehat{q}^+}$  & adjoint global mode vector \\
$\mathbf{R_i}$  & residual vector in subdomain $\Omega_i$ \\
$Re_c$  &  Reynolds number based on the chord length \\
$St$  &  Strouhal number \\
$T$ & temperature \\
$t$  & time \\
$u,v,w$  & velocities in x-,y- and z-directions respectively \\
$\mathbf{x}$  & vector of coordinate directions (x,y,z) \\
\end{longtable*}}

\section{Introduction}
\lettrine{T}{he} region behind a body with a blunt end, often referred as the base region, is normally characterized by having low momentum and low pressure, due to the flow separation that takes places at the end of the body\cite{Nash1963}. The flow detachment at this area is the prime contributor to pressure losses, playing a fundamental role in the aerodynamic design of the body. The afterbody of axisymmetric bodies\cite{Mariotti2018,Meliga2010}, the aft part of ramjets or aircrafts\cite{Thomas1985AircraftControl}, the rear flow of ground and flight vehicles\cite{Lamb1995}, or the trailing edge of turbine blades\cite{Denton1990,Wu2001} are, among others, fields where the base region flow drives the aerodynamic design process. 

\begin{figure}[t]
   \begin{minipage}{0.48\textwidth}
     \centering
     \includegraphics[height=5cm,keepaspectratio]{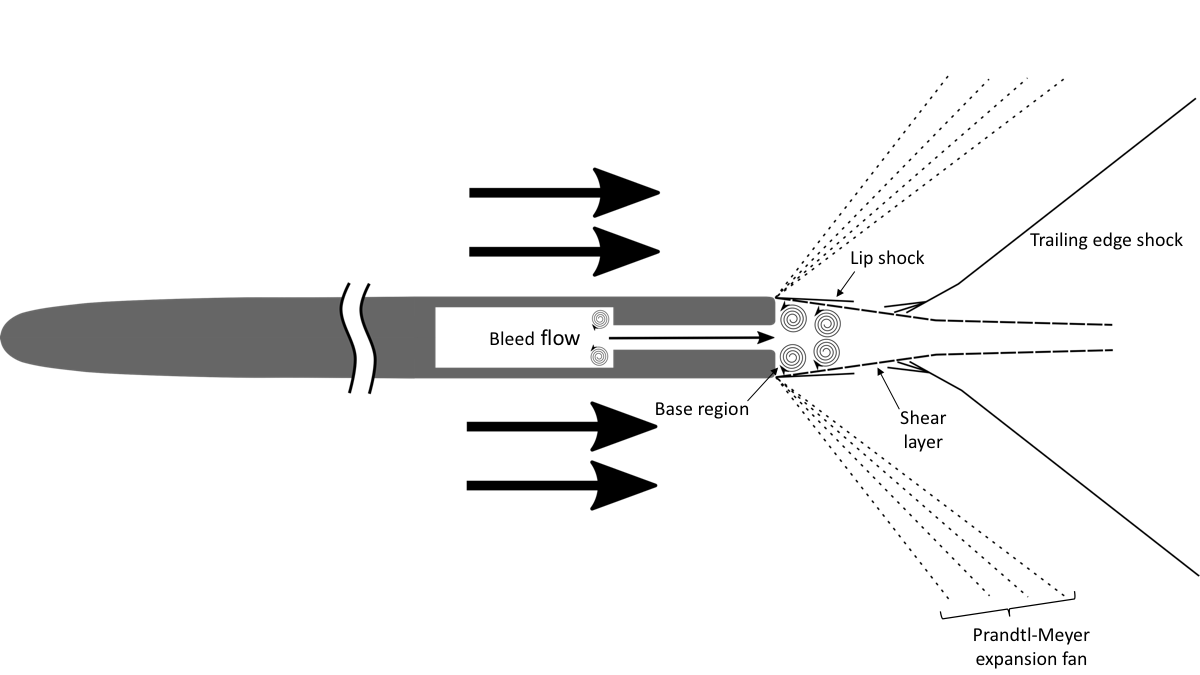}
     \caption{Flow topology at the base region in supersonic conditions}
	 \label{f:ftopology}
   \end{minipage}\hfill
   \begin {minipage}{0.48\textwidth}
     \centering
     \includegraphics[height=5cm,keepaspectratio]{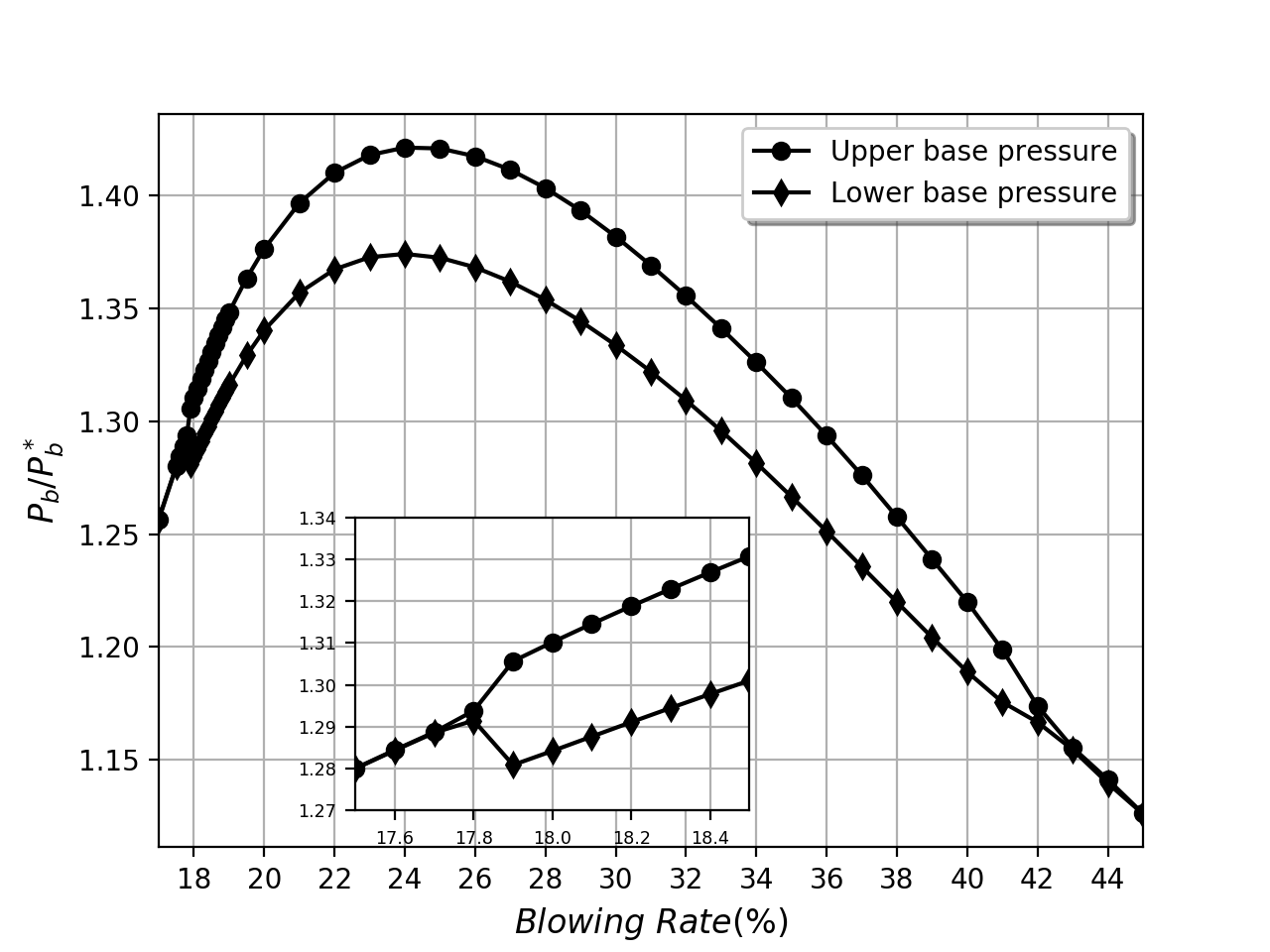}
	    \caption{Base region pressure bifurcation as a function of the bleed intensity. Figure extracted from \cite{Martinez-Cava2018}}
	    \label{f:pbifurcation}
   \end{minipage}
\end{figure}

The boundary layers from suction and pressure sides separate in an alternate manner, characterized by the Strouhal number, developing a pair of shear layers that lead to the formation of two (or more, depending on the Reynolds number) families of vortices. This unsteady flow separation is a source of tonal noise, also causing mechanical and thermal fatigue not only on the body of origin but on the ones situated downstream, affected by the traveling vortices and unsteady shock waves\cite{Paniagua2008UnsteadyAnalysis}. A schematic view of a two-dimensional base region in supersonic conditions is represented in Figure \ref{f:ftopology}. Different ways to control the base region properties have been considered, but one of particular interest is the so-called base bleed\cite{Motallebi1981, Tanner1975}, where flow is directly injected at the base region. This type of flow control is commonly found in turbomachinery, where colder flow from the turbine blade refrigeration circuits is ejected at the trailing edge of the thin airfoils, increasing the pressure at the base region and reducing the aerodynamic losses. Flow injection at the base region, even in small quantities, strongly affects the interaction between the two shear layers and modifies the flow topology in that region\cite{Saracoglu2013, Saavedra2017,Raffel1998}.


However, as it was found by Saracoglu et al.\cite{Saracoglu2013}, the injection of flow in the base region can lead to the apparition of non-symmetrical configurations related to a bifurcation phenomena associated to the intensity of the injected flow (Fig.\ref{f:pbifurcation}). Based on those results, Martinez-Cava et al.\cite{Martinez-Cava2018} performed a detailed study of a simplified turbine trailing edge combining RANS simulations and global stability analysis. Results showed that for jet intensities above certain value, a difference in pressure between the upper and lower sides of the trailing edge will appear, forcing the flow to change its direction through an induced Coanda effect. The stability analysis of this configuration identified the physical global mode responsible of the non-symmetrical conditions, its spatial structure, and its relationship with the blowing intensity. As a continuation of that study, an adjoint approach was used to identify the sensitivity areas of the instability\cite{Martinez-cava2018c} and its core regions, defined as those areas of the fluid domain where changes in the flow topology would have a major impact on the instability development. 

This study aims to fully characterize the transient flow topology around a base bleed configuration on a trailing edge, and the benefits that flow injected at the base region can have on flow control and aerodynamic performance. In particular, the main goal is to take advantage and promote the observed \textit{Coanda} effect, analyzing its capabilities and the physical mechanisms responsible of its generation. Despite of the abundance of literature on the study of flow bifurcation on a sudden expansion geometry\cite{Fearn1990, Alleborn1997FurtherExpansion, Battaglia2006BifurcationChannels}, only the references above mentioned cover the apparition of a non-symmetrical configuration on a base bleed configuration. One of the objectives of this research is therefore extend the available resources to identify this phenomenon and propose different ways to control and exploit it. To achieve this, a simplified body with a blunt trailing edge and a dedicated injection chamber for base bleed is considered, exploring a wide range of parameters to investigate their influence on the physical global modes responsible of the \textit{Coanda} effect. The sensitivity to changes with the Mach number, the base bleed intensity, and the shape of the trailing edge tip are explored, identifying those configurations with optimal characteristics for base flow control.

Flow solutions are obtained solving the compressible Navier-Stokes equations using an unsteady RANS approach, whereas the stability and sensitivity analysis will be computed using in house numerical libraries for large-scale eigenvalue problems. Furthermore, using the information gathered in the numerical analysis, passive and active flow control most sensitive regions are identified and characterized. Although it is not contemplated in this paper, this study goes beyond the numerical approach, using the same geometrical body to experimentally investigate the apparition of the \textit{Coanda} effect and correlate it with the numerical experiments. State-of-the-art flow visualization and measurement techniques are employed to analyze the flow around the body in a \textit{trisonic} wind tunnel facility. The numerical campaign is therefore designed to be validated by experimental means.

The rest of the paper is organized as follows. Section \ref{sec:Math-model} overviews the mathematical formulation, covering the aspects related with the flow solver, adjoint approach and the sensitivity fields calculations. Section \ref{sec:exp-met} summarizes the experimental setup and the description of the geometrical model, while in Section \ref{sec:num-exp} the numerical procedures to obtain the base flow and stability solutions are reviewed. In Section \ref{sec:results} and \ref{sec:sens-study} the main results of the flow characterization and flow control are described, finishing in Section \ref{sec:conclusions} with the main conclusions of this study.


\section{Mathematical model}\label{sec:Math-model}

\subsection{Flow solver} \label{sec:Flow-model}

The compressible version of the Reynolds Averaged Navier-Stokes equations (RANS) is used to model the flow, which set of equations can be written in conservative form as:

\begin{equation}\label{eq:NSvol}
\pdv{t} \int_\Omega \mathbf{q} \dd{\Omega} = -\int_{\partial \Omega} {\rttensor{F}} \cdot \mathbf{n} \dd{S},
\end{equation}

where vector\footnote{Bold variables are used to refer vectors.} $\mathbf{q}$ comprises the conservative variables (density, momentum and energy) and turbulent quantities, while $\Omega$ is a control volume with boundary $\partial \Omega$ and outer normal $\mathbf{n}$. $\rttensor{F}$ denotes the flux density tensor, which can be decomposed along the three Cartesian coordinate directions and  comprises the inviscid, viscous and turbulent fluxes, these last modeled with the SST version of Menter of the $k-\omega$ turbulence model\cite{Menter1994Two-EquationApplications}.

The flow domain $\Omega$ is discretized into a finite number of subdomains $\Omega_{i}, i=1 \ldots N$, where each subdomain contains $\mathcal{N}_f$ faces. For the calculation of steady solutions, the time-accurate two-dimensional Navier-Stokes equations are marched in time using a backward Euler implicit scheme, taking advantage of local time stepping and multigrid algorithms to accelerate convergence.

The spatial discretization of system (\ref{eq:NSvol}) gives rise to a system of ordinary differential equations, that can be written in general form for a subdomain $\Omega_{i}$ as:

\begin{equation}\label{eq:dW_R}
{|\Omega|_i} \pdv{\mathbf{q}_i}{t} + \mathbf{R}_i = \mathbf{0}, \qquad \mathbf{R}_i = \sum_{j=1}^{\mathcal{N}_f} {\rttensor{F}}_j \mathbf{n}_j, \qquad i=1 \ldots N
\end{equation}

where $\mathbf{R}_i$ is equivalent to the flux contributions, and $\mathbf{q}_i$ represents a discretized vector state solution. Both vectors have dimensions $N_{v}$, that depend on the number of fluid variables considered. In this scenario, with a two-dimensional flow solution using the $k-\omega$ turbulence model, $N_v=6$, with $\mathbf{q}_i = \mathbf{q}_i(\rho,\rho u, \rho w, \rho E, \rho k, \rho \omega)$.

Non-slip and adiabatic wall boundary conditions are imposed on the body surface as:

\begin{equation}\label{eq:BC}
u=w=0  \qquad  \pdv{T}{\mathbf{n}}=0
\end{equation}

where $\mathbf{n}$ is the normal direction to the body surface and $T$ stands for the temperature. 


\subsection{Stability analysis} \label{sec:stab_theory}

Stability analysis studies the growth or decay of perturbations superimposed onto a (usually) steady solution of the Navier-Stokes (NS) equations. The analysis can identify which particular features are prone to evolve under slight modifications of the flow conditions, either by introducing a perturbation or caused by a modification of some physical or geometrical parameters. Examples of the application of stability and sensitivity analysis to fluid dynamics problems can be found in the literature for a large variety of flow topologies \cite{Theofilis2011,Luchini2014,Chomaz2005}.

If the flow solution was obtained as a steady point of equilibrium of Eq. (\ref{eq:dW_R}), $\mathbf{R}(\mathbf{\bar{q}_b})=0$ and $\mathbf{\bar{q}_b}$ will be denoted as \textit{base flow}. On the contrary, if the flow solution is non-stationary and periodic, with $\mathbf{q}(\mathbf{x},t) = \mathbf{q}(\mathbf{x},t+T)$ a \textit{mean flow} $\mathbf{\bar{q}}$ can be obtained by averaging the unsteady RANS (URANS) solution\cite{Barkley2006,Beneddine2016ConditionsAnalysis}. 
In both cases (subindex \textit{b} is dropped for simplicity), the flow solution can be decomposed into a non-time-dependant term and small fluctuations fields:
\begin{equation}\label{eq:FV5}
\mathbf{q}(\mathbf{x},t) = \mathbf{\bar{q}}(\mathbf{x}) + \varepsilon \mathbf{\widetilde{q}}(\mathbf{x},t),
\end{equation}
with $\varepsilon\ll 1$. This last term will be the subject of the study on the stability analysis, to evaluate if those fluctuations are prone to evolve and become dominant.




The separability of temporal and spatial derivatives in (\ref{eq:FV5}) permits the introduction of an explicit harmonic temporal dependence of the disturbance quantities. Moreover, in pure two dimensional cases the velocity component and all derivatives in the spanwise (y) direction are neglected, and solutions can be assumed in the form of:

\begin{equation}\label{eq:ansatz}
	\mathbf{\widetilde{q}}(\mathbf{x},t) = \mathbf{\widehat{q}}(x,z)e^{\sigma t},
\end{equation}
where $\sigma$ is a complex scalar and the complex vector $\mathbf{\widehat{q}}$ describes the spatial complex disturbance. Introducing (\ref{eq:ansatz}) and (\ref{eq:FV5}) into system (\ref{eq:dW_R}), a generalized eigenvalue problem is recovered:

\begin{equation}\label{eq:GEV}
    \mathbf{A}\widehat{\mathbf{q}}=\sigma \mathbf{M} \widehat{\mathbf{q}}.
\end{equation}

The diagonal matrix $\mathbf{M}$, with leading dimension $N_{v} \times N$, contains the volumes associated to each cell, $\sigma=\sigma_r+i\sigma_i$ is the complex eigenvalue, and $ \mathbf{A}=\left[\pdv*{\mathbf{R}}{\mathbf{q}}\right]_{\mathbf{\bar{q}}}$ represents the linearized jacobian matrix of the fluxes evaluated in the base flow.with $\mathbf{A}=\qty[\pdv*{\mathbf{R}}{\mathbf{q}}]_{\mathbf{\bar{q}}}$. The real part of the eigenvalue, $\sigma_r$, refers to the amplification rate of the corresponding eigenmode, with the imaginary part $\sigma_i$ being the pulsation, related to the associated frequency through the Strouhal number, as $St = 2\pi / \sigma_i$. 


\subsection{Adjoint eigenvalue problem}
We first define the following discrete inner product for an arbitrary pair of complex vectors, $\mathbf{p}$ and $\mathbf{q}$:
\begin{equation}
	< \mathbf{p},\mathbf{q} > = 
			\int_\Omega \mathbf{p}^H \mathbf{q} \dd{\Omega} = \mathbf{p}^H \mathbf{M q }
\end{equation}
where \textit{H} denotes the conjugate transpose and $\mathbf{M}$ is the diagonal and invertible volumes matrix previously defined.

Following this, the matrix $\mathbf{A}$ and its discrete adjoint operator $\mathbf{A}^+$ can be easily related as $\mathbf{A}^+ = \mathbf{M^{-1}} \mathbf{A}^H \mathbf{M}$, allowing to obtain the associated adjoint eigenvalues, $\sigma^+$, and eigenmodes, $\mathbf{q}^+$:
\begin{equation}\label{eq:GEV2-ADJ}
    \mathbf{A}^+\widehat{\mathbf{q}}^+ = \sigma^+ \mathbf{M} \widehat{\mathbf{q}}^+
\end{equation}


\subsection{Sensitivity analysis}
The information contained on the adjoint eigenmodes can be used to evaluate the sensitivity of the eigenvalues of the system. Giannetti and Luchini\cite{Giannetti2007} recovered the concept of the 'wavemaker' for two-dimensional flows, showing that the areas of the flow that were more sensitive to structural changes (a localized force-velocity coupling) were localized on the overlapping regions of the direct and adjoint modes. This defines the structural sensitivity as:
\begin{equation}
	\nabla_s \sigma = \frac{\| \widehat{\mathbf{q}}^+ \| \| \widehat{\mathbf{q}} \| }{< \widehat{\mathbf{q}}^+ , \widehat{\mathbf{q}}> }
\end{equation}
where $<\bullet,\bullet>$ denotes the discrete inner product, previously defined, and its associated norm $\parallel \bullet \parallel = <\bullet,\bullet>^{1/2}$. In this work, the adjoint modes have been normalized according to the condition $< \widehat{\mathbf{q}}^+ , \widehat{\mathbf{q}}>=1$.

Marquet et al.\cite{Marquet2008} studied the effects on the eigenvalue drift caused by a perturbation acting as a base flow modification or as a source forcing term. This context was developed afterwards in a discrete framework \cite{Browne2014, Mettot2014}, so is briefly described here.
The sensitivity of a specific eigenvalue to base flow modifications is a complex vector field, that can be calculated as the product of the adjoint operator of the sensitivity matrix $\mathbf{B}(\mathbf{\bar{q},\widehat{q}})$ and the associated eigenmode:
\begin{equation}
	\label{eq:sensbflow}
	\nabla_{\mathbf{\bar{q}}} \sigma = \qty(\pdv{\mathbf{A}(\mathbf{\bar{q})\widehat{q}}}{\widehat{\mathbf{q}}})^+ \widehat{\mathbf{q}}^+ \ = \ \mathbf{B}^+(\mathbf{\bar{q},\widehat{q}}) \widehat{\mathbf{q}}^+
\end{equation}

The matrix operator $\mathbf{B}(\mathbf{\bar{q},\widehat{q}})$ is obtained through differentiating the jacobian matrix and the direct mode related to the eigenvalue of interest.
This methodology can be extended to the application of a steady force as an external perturbation, $\delta \mathbf{q_f}$, and the respective sensitivity field is obtained:
\begin{equation}
	\label{eq:sensforcing}
	\nabla_{\mathbf{q_f}} \sigma = \mathbf{\bar{q}}^+ = (\mathbf{A}^+(\mathbf{\bar{q}}))^{-1} \nabla_{\mathbf{\bar{q}}} \sigma
\end{equation}
being $\mathbf{\bar{q}}^+$ the adjoint operator of the mean flow, obtained through the solving of the linear system (\ref{eq:sensforcing}).


\section{Experimental methodology and model description} \label{sec:exp-met}
Although no experimental results are provided in this paper, the experimental analysis is a crucial part of this research. For illustration, and for a better understanding of the numerical experiments design, the experimental set up is briefly described here.

The Purdue Experimental Turbine Aerothermal Lab (PETAL) facility has a linear wind tunnel conceived for low Technology Readiness Level (TRL 1-2) studies, which perfectly matches the requirements for the experiments to be carried in this research. It is a \textit{trisonic} facility, covering from low subsonic to supersonic conditions, together with a wide envelope of Reynolds number conditions. This is accomplished through the independent control of inlet total pressure and temperature with a heat exchanger, as well as the control of the downstream pressure from atmospheric to vacuum conditions.
The test section has a prismatic shape with dimensions 540x230x170 mm (length x width x height), and its unique visual access from the 4 fully transparent windows allow to combine different flow visualization techniques to characterize the aerothermal phenomena involved in the problem treated in this research. The PETAL facilities allow to use optical measurement techniques such as high frequency particle image velocimetry (PIV) and Schlieren flow visualization, that will be combined with pressure and heat flux measurements to provide an extended dataset of the experimental analysis. An extended description of the PETAL facilities can be consulted on \cite{Paniagua2016DesignMeasurements}.

\begin{figure}
    \centering
    \includegraphics[width=0.2\textwidth]{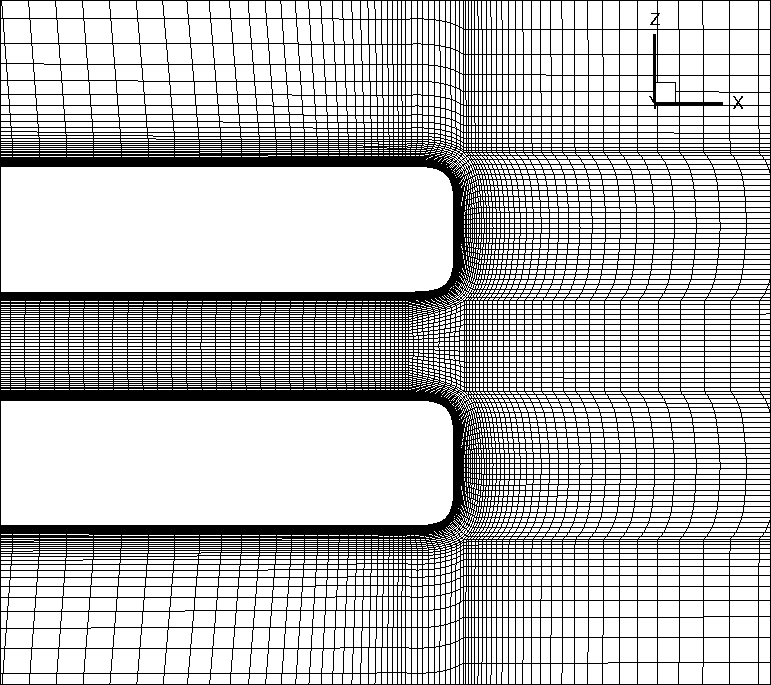}
    \includegraphics[width=0.2\textwidth]{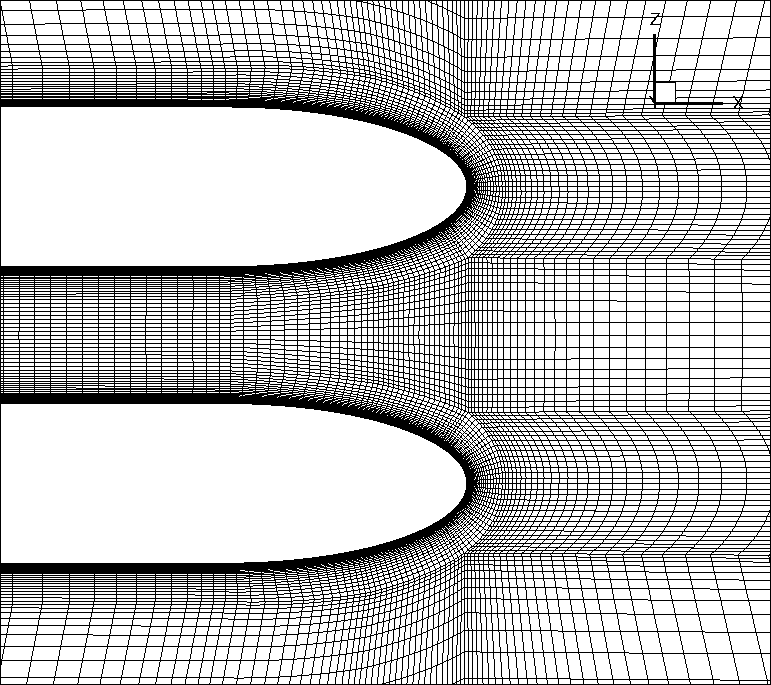}
    \caption{Comparison of trailing edge shapes used on the analysis. Left image shows the ``straight" (super ellipse) shape, with the ``rounded" (ellipse) shape on the right image.}
    \label{fig:TE-shapes}
\end{figure}

The geometry of the model was therefore designed to be used in the \textit{trisonic} range, with a blockage ratio with respect to the tunnel height lower than 8 to avoid excessive interference of the wind tunnel walls with the flow topology. The body consisted on a zero-camber airfoil of 160 mm chord (\textit{c}), with a fitness ratio of 8, where a Haack Series\cite{Perkins} nose shape was chosen to adapt the flow without any detachments or very strong gradients for the three investigated flow regimes. The base bleed injection system was defined by a plenum area where the cooling flow is injected in subsonic conditions, and a straight channel into which the flow passes through a contraction that is finally discharged at the base region. The aspect ratio of the contraction was kept to 3, to ensure subsonic conditions at the plenum area, ending on a base bleed slot of 6 mm height. Two trailing edge tips were considered for the analysis, to study the variability of the \textit{Coanda} effect with the geometry shape. An almost straight shape, defined by a supper-ellipse with an exponential parameter of 5, was used as a reference case, and will be referred as ``straight" trailing edge along this paper. This geometry was similar to the one studied in previous works\cite{Saracoglu2013, Martinez-Cava2018, Martinez-cava2018c}, where a mild-\textit{Coanda} effect (where the flow was deflected by a pressure difference and will follow the shape of a recirculation area, but without been attached to a surface) was observed to appear at the base region, due to the first instability of the sudden expansion of the base bleed channel. The second trailing edge considered for analysis was designed with an ellipsoidal shape, where its curvature was expected to produce a full \textit{Coanda} effect (with a flow deflection related to the attached flow following the curvature of just one of the surfaces). This shape will be referred as ``rounded" along this paper. Both trailing geometries are shown on Figure \ref{fig:TE-shapes}. 

\section{Numerical experiments} \label{sec:num-exp}
To predict the behavior of the model and to perform a design of experiments, the geometry of the two-dimensional model and the wind tunnel walls were reproduced on a set of numerical simulations. To avoid non-physical boundaries on the computational domain, further inflow and outflow boundaries were imposed on those simulations were subsonic conditions were present.

\begin{figure}
    \centering
    \includegraphics[width=0.95\textwidth]{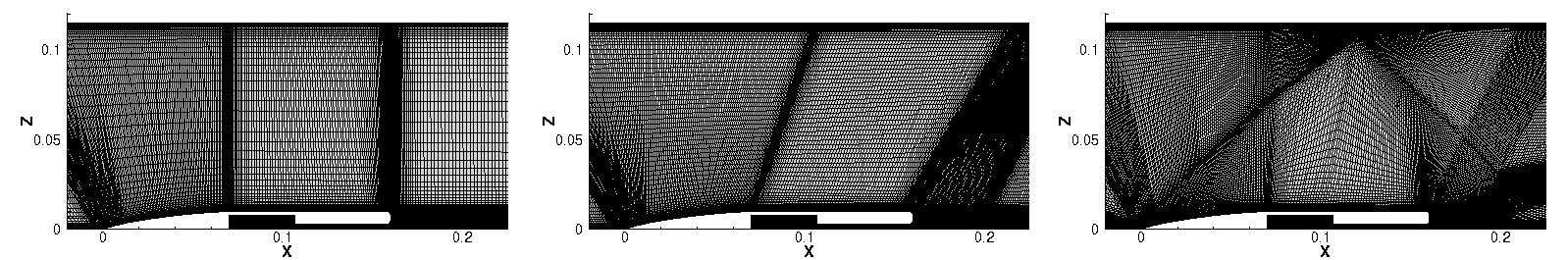}
    \caption{Mesh topology used for the different sonic regimes: Subsonic (left), transonic (centre), supersonic (right).}
    \label{fig:mesh-topology}
\end{figure}

Three different sonic regimes were investigated, with an optimized mesh topology for each configuration (Fig. \ref{fig:mesh-topology}). A Mach number at the inflow of 0.4 was taken to analyze the behavior of the cooling flow at subsonic conditions; an inflow Mach number of 0.7, that produced a flow speed of Mach=1.1 at the trailing edge of the model, was chosen to evaluate transonic conditions; and a third configuration with flow speed inflow of Mach=2 was used for supersonic flows analysis.

\begin{figure}
    \centering
    \includegraphics[width=0.7\textwidth]{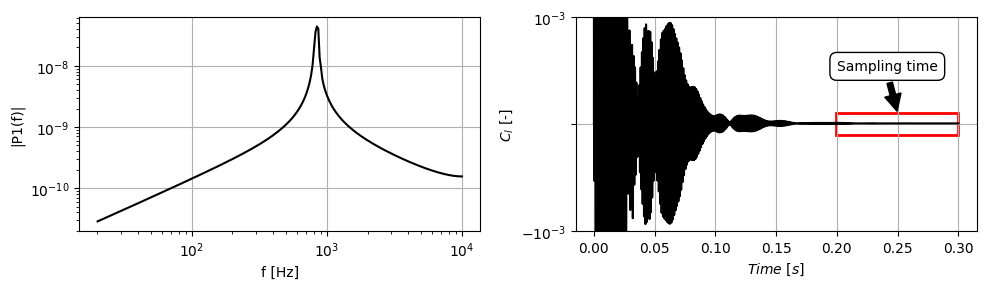}
    \caption{FFT analysis (left) of the lift coefficient sampling from an unsteady subsonic simulation (right).}
    \label{fig:muestreo-f}
\end{figure}

Grid spatial convergence was evaluated for a particular time step of the URANS simulations, with five different meshes (M0, M1, M2, M3, and M4, with each following mesh having double the points of the previous one) being used for the analysis. The Grid Convergence Index (GCI)\cite{Zhou2018Grid-convergedTube} was evaluated for three meshes, keeping the coarsest mesh (M0) as the stencil for the convergence analysis. The density values of each mesh were interpolated into M0, and the GCI was calculated using the Root Mean Square values. The results of the analysis are not included here for conciseness. Temporal discretization was done ensuring that each period of the dominant temporal fluctuations was covered with at least 40 timesteps. Due to this criteria, different time steps were used for each regime, varying from $10^{-5}$ s for subsonic conditions, to $10^{-6}$ s for supersonic flow conditions. The values of the monitored flow variables were extracted or averaged in a window of at least 100 periods, after the flow was already developed into a periodic state (Fig. \ref{fig:muestreo-f}).

On the subsonic and transonic meshes, total pressure and total temperature were imposed at the inlet, with an ambient pressure at the outlet boundary. On the mesh used for supersonic conditions, constant values were imposed at the inlet using a supersonic inflow boundary condition, with an exit-pressure condition at the outlet, were a differentiation of direct extrapolation or back pressure condition was done depending on the flow Mach number at the outlet cells. For all the meshes, adiabatic viscous walls were considered for the upper and lower walls, and for the blade model. Finally, a reservoir-pressure inflow was used to provide the cooling flow through the plenum area, imposing total pressure and temperature values. For the sake of readability, the coefficient $C_b = p_{0purge}/p_{0\infty}$, relating the total pressure of the reservoir pressure boundary condition with the total free stream pressure, was chosen to describe the pressure ratio of the base bleed flow.

A global stability analysis of the flow solutions obtained through RANS and URANS solutions requires the generation and storage of the linearized Jacobian squared matrix. The real non-symmetric operator $\mathbf{A}$ (Eq. \ref{eq:GEV}) was extracted using the first-discretize-then-linearize technique implemented in the DLR TAU-code solver from either \textit{base} or \textit{mean} flows, and saved using a compressed sparse row format. The generalized eigenvalue problem was solved using the UPM in-house numerical library for large eigenvalue problems, zTAUev, which has been validated for different configurations\cite{Iorio2014,Browne2014,Gonzalez2017}. Our tool exploits the capabilities of external libraries (as PETSc\cite{petsc-efficient}, MUMPS\cite{Amestoy2001AScheduling} or ARPACK\cite{Lehoucq1998ARPACKGuide}) to perform a complete Lower-Upper (LU) decomposition of the input matrix and obtain the eigenvalues of interests, being these only a small range of the eigenvalues of the total spectra, using the Implicitly Restarted Arnoldi Method (IRAM) algorithm. Likewise, direct and adjoint eigenmodes of the system were extracted and processed afterwards to obtain the sensitivity maps.

Since the study aims to understand the apparition of the \textit{Coanda} effect for different flow configurations, the nature and development of the non-symmetrical behavior of the injected flow needs to be analyzed from its onset, this implying the need of doing a global stability analysis around a symmetric base flow. To enforce this symmetry, only half of the domain was considered for the numerical simulations performed for stability analysis, imposing a symmetry boundary condition on the plane of symmetry (\textit{xy}). After obtaining a ``half" domain flow solution, mesh and solution were mirrored into the symmetry plane and the stability analysis of the full domain was carried out. The global analysis of the mirrored-full domain around this symmetric flow at different flow conditions will show non-symmetric modes in the analysis, which could eventually grow and generate a new non-symmetric flow configuration.


\section{Flow characterization} \label{sec:results} 
\subsection{Subsonic flow}
\begin{figure}
    \centering
    \includegraphics[width=0.95\textwidth]{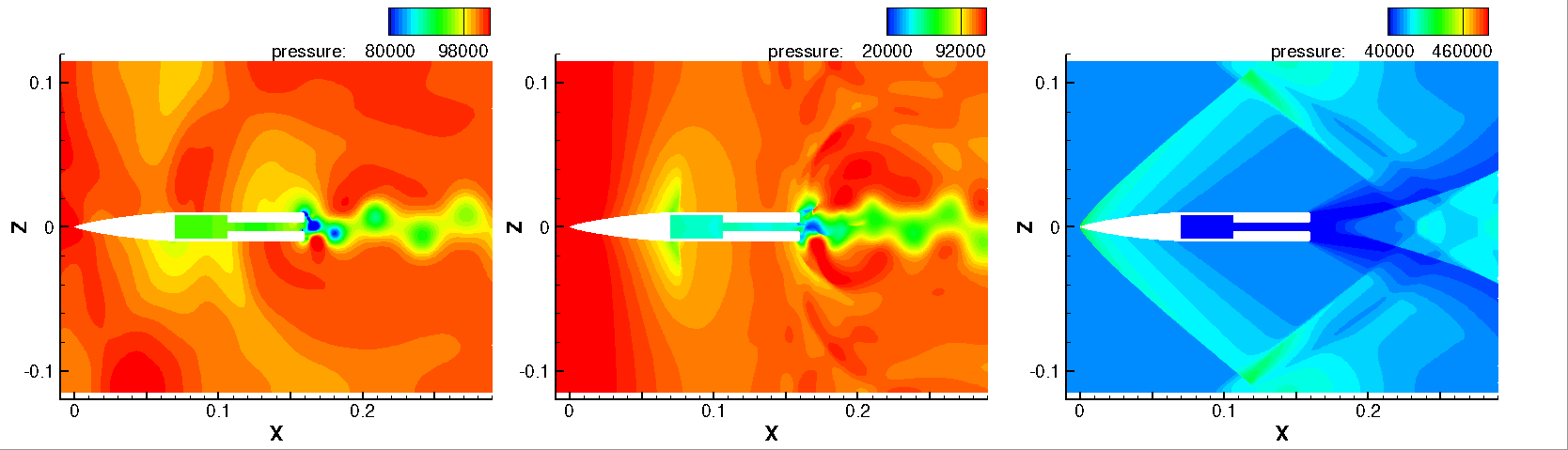}
    \caption{Temporal snapshot of static pressure contours for subsonic (left), transonic (center), and supersonic (right) regimes. Base bleed not active.}
    \label{fig:baseflows}
\end{figure}

Subsonic analysis was designed for an inflow condition of Mach=0.4, $Re_c=9.5\times 10^5$, total free stream temperature of 430K, and with a static temperature for the base bleed of 300K. The flow behind the body is characterized by the classic Von-Karman vortex street, with a characteristic Strouhal number of 0.264. Vortices separate from the upper and lower sides in an alternate manner, producing an oscillating behavior of the wake and the base region (Fig. \ref{fig:baseflows}-left). The two shapes of the tested trailing edge have little influence on the flow topology when no cooling flow is applied at the base region, showing differences on the order of 0.1 KHz on the frequency of the vortex shedding.

\begin{figure}
    \centering
    \includegraphics[width=0.99\textwidth]{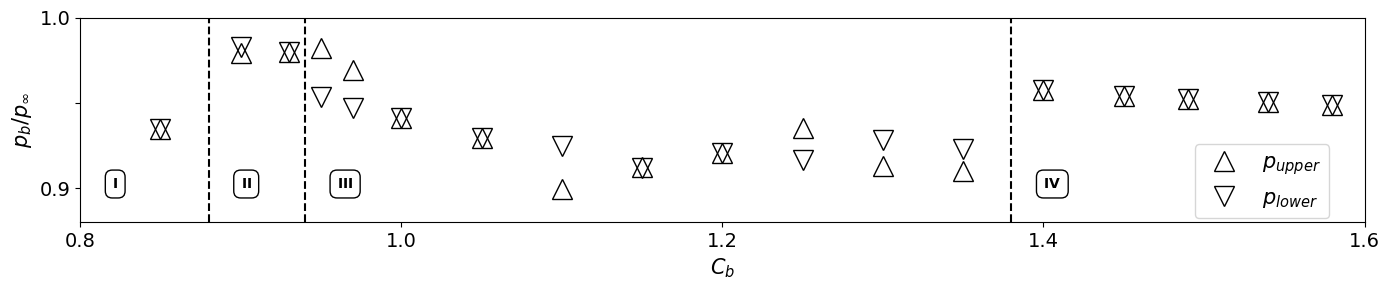}
    \includegraphics[width=0.99\textwidth]{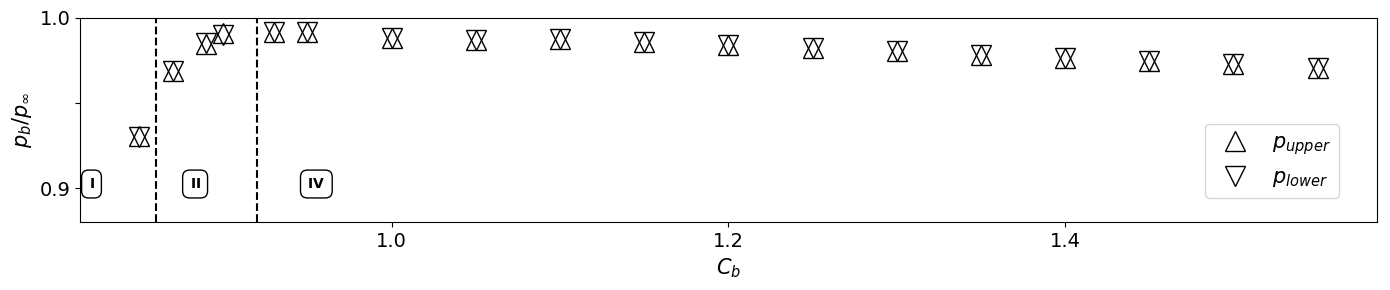}
    \caption{Base pressure evolution with the base bleed intensity in subsonic regime. Results for straight (upper) and rounded (lower) trailing edge geometries.}
    \label{fig:pb-sub}
\end{figure}

The dimensionless pressure relations for straight and rounded trailing edges when base bleed is applied can be seen in Fig. \ref{fig:pb-sub}, where the base pressure of upper and lower sides of the trailing edge are plotted against the base bleed intensity. The effects produced by the base bleed flow on a straight trailing edge can be divided in four different phases, denoted on the figure from phase I to phase IV. First, when cooling flow is ejected with low intensity at the base region, an initial filling effect is produced and the base pressure increases. After that, the bleed flow introduces a source of symmetry on the wake downstream, and the vortex shedding is actually neutralized for a certain range of blowing rates. It is at these rates when the first presence of the non-symmetrical flow can be detected, as the first instability of the sudden expansion appears. Named as phase II, is at this stage when the base bleed flow is deflected towards either the upper or lower side of the trailing edge in a mild \textit{Coanda} effect style, as the flow is attached to the recirculation areas at the base region. However, when the pressure ratio $C_b > 0.95$, the presence of the vortex shedding reappears with a frequency that is twice the one present with no blowing. This increment in frequency has been previosly related in the literature\cite{Motallebi1981}, and the authors believe that this happens due to the onset of the secondary instability of the geometrical sudden expansion of the base bleed jet channel. Base pressure decays at these blowing rates, reaching values similar to those without blowing, and with constant asymmetries on the pressure averaged values of upper and lower sides, showing an change in direction of the ejected flow that is not periodic, but strongly affected by the secondary vortices separated from the injection channel. Finally, at phase IV, for base bleed values with $C_b > 1.4$ the flow inside the injection channel becomes supersonic and the momentum of the ejected flow keeps the flow symmetric, weakening any oscillation and increasing the base pressure towards a final plateau of $p_b/p_{\infty} \approx 0.94$.

If the rounded trailing edge geometry is considered, the effects of the base bleed are milder, as the geometrical sudden expansion is not present. First, an increment on the base pressure is observed as the base region is filled with cooler flow. For values of $C_b$ as low as 0.87, the non-symmetrical flow configuration of phase II already appears, as a full \textit{Coanda} effect can be observed at the end of the injection channel. However, the effect quickly disappears, and for base bleed intensities higher than $C_b \approx 0.92$ the base pressure remains constant, without any vortex presence or flow oscillation downstream the trailing edge. Phase III, related with reapparition of the shedding with higher frequency, is not present in this case.

\begin{figure}
    \centering
    \includegraphics[width=0.95\textwidth]{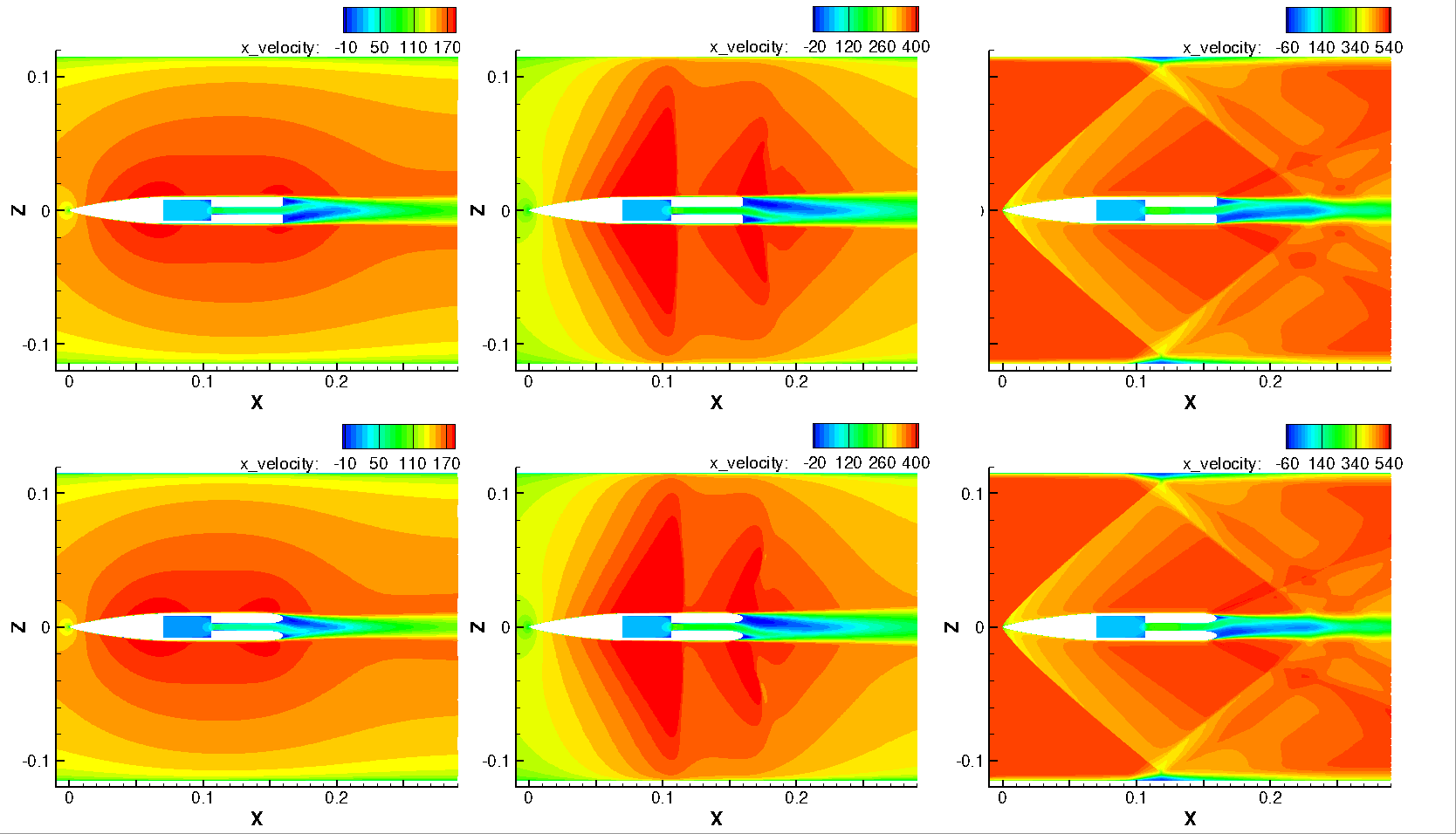}
    \caption{Temporal snapthot of streamwise velocity contours for base bleed non-symmetrical flow configurations (phase II), caused by the global instability of the sudden expansion. Upper row shows straight trailing edge flow solutions (subsonic to supersonic, from left to right), and bottom row shows the rounded trailing edge flow solutions.}
    \label{fig:non-symmetry}
\end{figure}

Whereas the non-symmetrical configurations with a straight trailing edge are related with a difference in pressure between the upper and lower sides, as the mild-\textit{Coanda} effect takes place at the base region when one of the recirculation areas (upper/lower) grows at the expense of the contrary, the apparition of a non-symmetrical configuration on a rounded trailing edge is not linked with a pressure bifurcation. The identification of this phenomenon is therefore linked to flow visualization or velocity components monitoring. Velocity contours are shown on Figure \ref{fig:non-symmetry}, for a better illustration.

\begin{figure}
    \centering
    \includegraphics[width=0.95\textwidth]{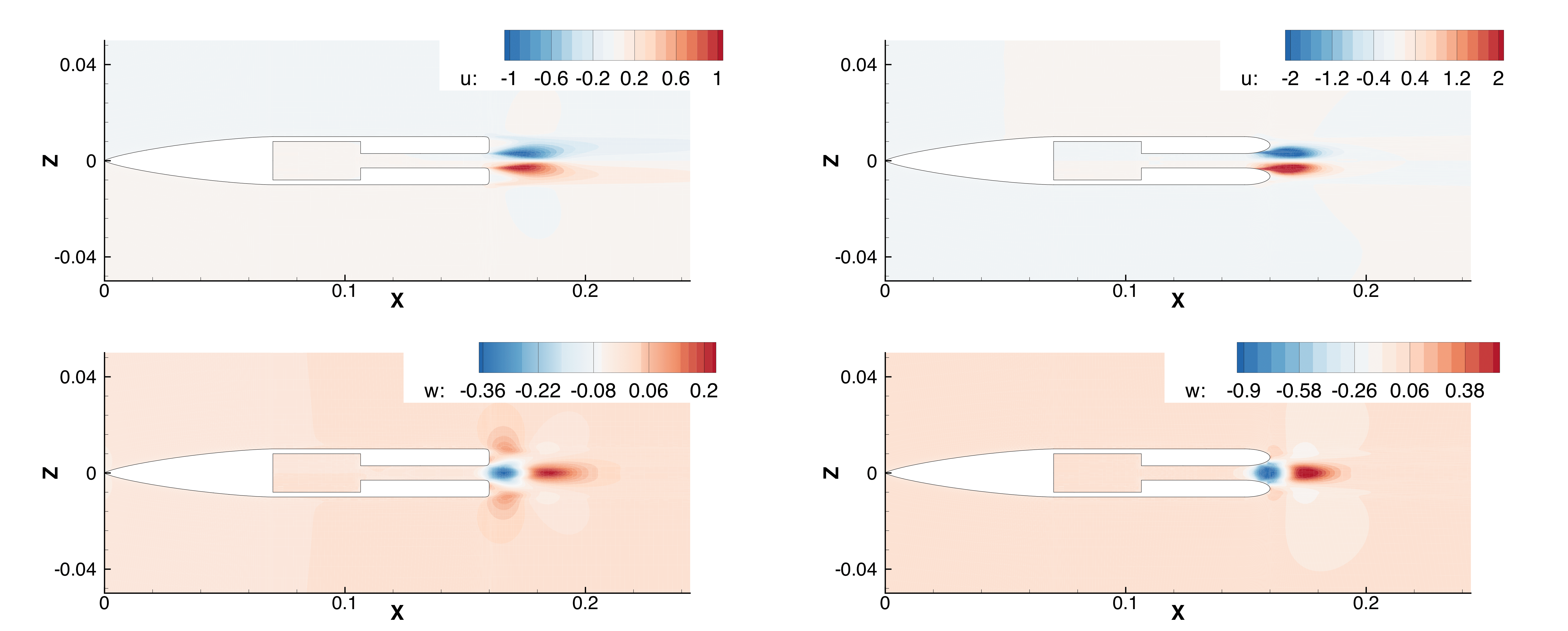}
    \caption{Global mode related with the geometrical sudden expansion, subsonic regime. Streamwise (\textit{u}) and vertical (\textit{w}) velocity perturbations are shown for both trailing edge configurations.}
    \label{fig:sub-globalmode}
\end{figure}

The stability analysis of configurations from phase II (non-symmetrical flow condition) allows to recover the global mode related with the sudden expansion of the channel, for both trailing edge geometries. For only few configurations a converged steady RANS solution (\textit{base region} approach) could be obtained. As the sources of unsteadiness in the flow are multiple, a \textit{mean flow} analysis was therefore performed. 300 periods of oscillations from a URANS simulation over half domain (cropped in the \textit{xy plane}) were averaged to extract the \textit{mean flow}. Streamwise and vertical velocity perturbations contours of the global mode related with the instability are shown in Figure \ref{fig:sub-globalmode}. The shape of the eigenmode is similar to the identified in a sudden expansion phenomena, with the main non-symmetrical streamwise perturbations located at the base region in the form of two anti-symmetrical lobes. This global mode becomes unstable for a certain range of blowing rates and drives the change of direction of the ejected flow\cite{Martinez-Cava2018}. The eigenvalue associated with the sudden expansion global mode has no associated frequency, despite the pressence of oscillations in the flow, these related with the vortex shedding. The instability is stable in time, and the flow will remain attached the side it initially contacted, even if an oscillating vortex shedding co-exists downstream.

\subsection{Transonic flow}
Inflow total temperature was kept constant at 430K, with a Reynolds number of approximately $Re_c$=$1.8 \times 10^6$, and a Mach number of 0.7. The static temperature for the base bleed was again set constant to 300K. The flow accelerates over the model, reaching the trailing edge at a speed of Mach=1.1. Two symmetric shock waves develop at the inflection point of the geometry, at the end of the nose ogive, which position will remain almost constant, slightly varying for each base bleed intensity. The blockage effect of the tunnel walls produces an ``unstart" effect on the downstream shock waves, as the oscillation of the wake behind the body limits the mass flow that passes through the trailing edge section (Fig. \ref{fig:baseflows}-center). The trailing edge shock waves appear downstream the body and travel downstream with the separated vortices in an alternate manner with an associated Strouhal number of 0.324.

\begin{figure}
    \centering
    \includegraphics[width=0.99\textwidth]{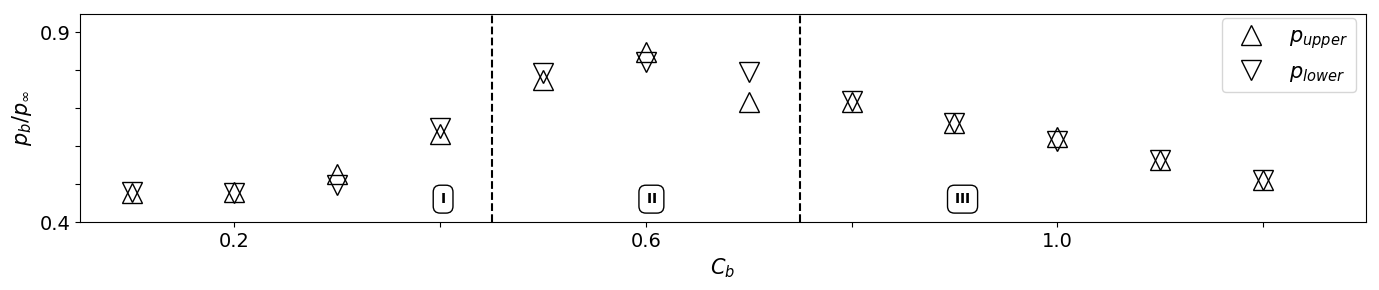}
    \includegraphics[width=0.99\textwidth]{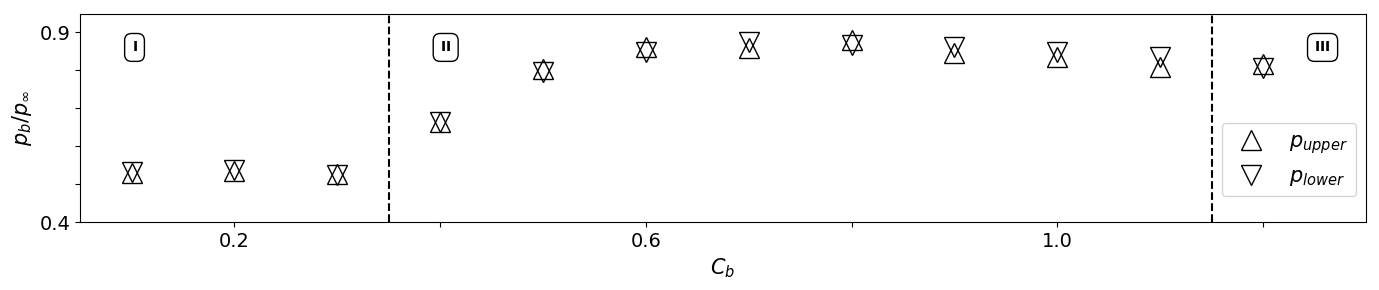}
    \caption{Base pressure evolution with the base bleed intensity in transonic regime. Results for straight (upper) and rounded (lower) trailing edge geometries.}
    \label{fig:pb-trans}
\end{figure}

Interestingly, for this regime the shape of the trailing edge shows to have a major impact on the base region flow behavior (Fig. \ref{fig:pb-trans}). If the straight trailing edge tip is used, the application of base bleed on this configuration produces again an initial increase in base pressure (phase I), followed by an apparition of a non-symmetric condition for those blowing rates related with maximum base pressure (phase II). This stage is linked with a strong reduction in the vortex shedding frequency, which eventually increases again with the final decay in base pressure as the base bleed flow speeds reach supersonic regime (phase III). However, a rounded trailing edge shape shows a delay on the later decay in pressure, at the same time that the \textit{Coanda} effect prevails over the symmetric condition for larger blowing rates.

The stability analysis to recover the global mode related with the Pitchfork bifurcation of these configurations could however not be completed with the current methodology. Due to the nature of the flow topology, the solutions of a simulation of a mirrored half domain (to ensure symmetric conditions) will not match the complementary flow solution of a full domain simulation, an essential requirement for the methodology here presented. Other methodologies (continuation methods, for example) will be applied in the future to tackle this problem.

\subsection{Supersonic flow}\label{sec:flow-sup}
Due to the isentropic expansion of the flow that occurs through the convergent-divergent geometry that would be installed at the wind tunnel to analyze supersonic conditions, the total temperature at the inflow boundary considered for the supersonic flow simulations was decreased to 300K. To still provide cooler flow at the base bleed, the temperature of the injected flow was corrected to 120K. Inflow speed was set at Mach=2.0 at the boundary. The flow around the body is purely supersonic, except for the base region area where a subsonic recirculation section takes place, confined by the strong shear layers of upper and lower sides of the trailing edge. Two oblique shock waves depart from the sharp leading edge, being reflected at the upper and lower walls and impacting downstream the base region with the trailing edge shock system (Fig. \ref{fig:baseflows}-right). This flow system has been carefully studied in the past\cite{Martinez-Cava2018,Saracoglu2013}, so it is briefly described here. As the boundary layers of upper and lower sides approach the trailing edge, they separate forming two symmetric shear layers that confine a subsonic recirculation area, the base region. The main flow first expands and accelerates through a Prandtl-Meyer expansion fan, just to later compress and adapt through a strong trailing edge shock wave. When no base bleed is applied, and there is no pressure difference between the upper and lower sides of the trailing edge, the flow downstream remains symmetric and steady, with no vortex shedding present.

The effects of base bleed on supersonic flow were described in detail by Saracoglu et al.\cite{Saracoglu2013} for an inflow Mach number of 1.5 and a straight trailing edge, and few changes from what was described in that scenario were found here. We again divide the flow topology changes in four phases, as the blowing rate increases. Low intensities are gathered in phase I, where base pressure increases keeping the symmetry of the flow. At a bleed intensity of $C_b=0.1$, a pressure difference appears between upper and lower sides, indicating the appearance of the sudden expansion first bifurcation. We marked this stage as phase II, that lasts up to $C_b=0.16$ and covers the maximum values of base pressure that are produced by the flow injection. Higher mass flows produce a decrement on the base pressure as the wake is pushed downstream by the jet and the base region flow is entrained into the base bleed flow. For values of $C_b$ higher than 0.20, strong frequency fluctuations of the order of $St=0.388$ appears, showing a non-symmetrical configuration with strong wake oscillations for $C_b=0.25$. This is linked with an increase in base pressure, that again drops when the symmetric state is recovered. Finally, phase IV arrives for $C_b$ values of 0.28 and higher, where the intensity of the base bleed flow keeps the flow symmetric with base pressure values similar to those without base bleed.

When a rounded trailing edge is considered, the \textit{Coanda} effect appears early, as the flow is rapidly attached to one of the sides of the base bleed channel. A jet flow deflection can be observed for values of $C_b$ between 0.08 and 0.22, with a maximum base pressure for a mass flow of $C_b=0.14$. At higher blowing rates it starts to decay. The strength of the jet eventually recovers the symmetric state, but without the appearance of high frequency oscillations, as it was shown for the straight trailing edge. Flow solutions for non-symmetrical configurations are shown on Figure \ref{fig:non-symmetry}.

\begin{figure}
    \centering
    \includegraphics[width=0.99\textwidth]{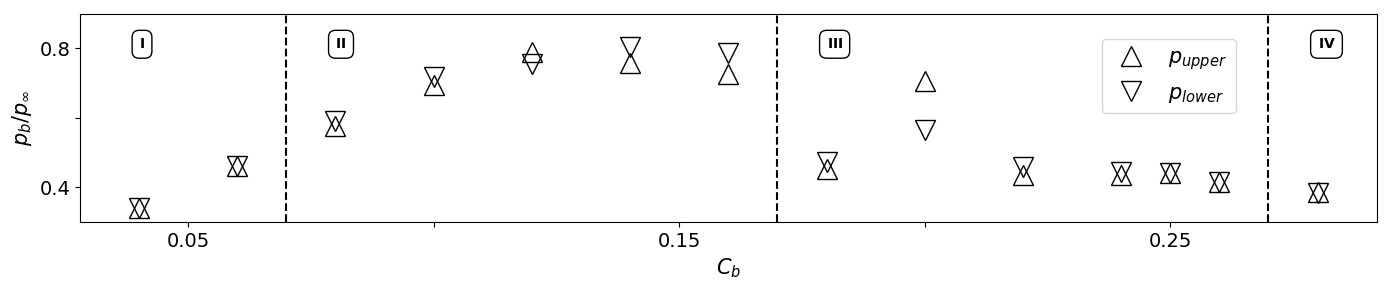}
    \includegraphics[width=0.99\textwidth]{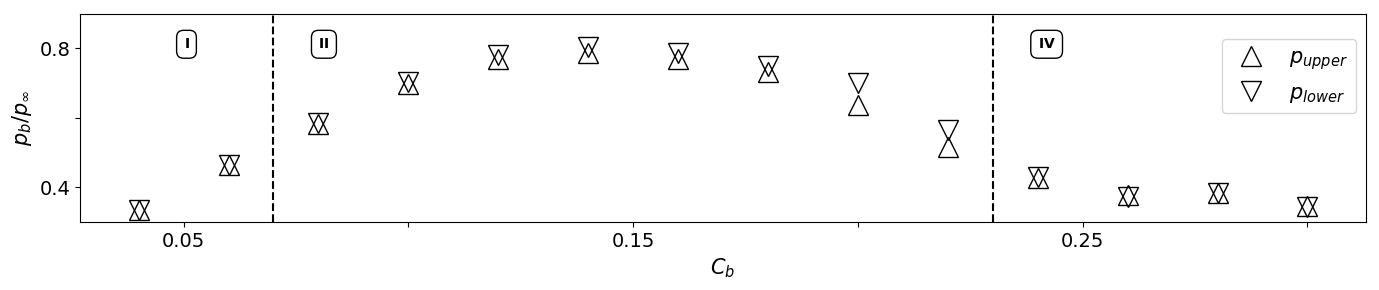}
    \caption{Base pressure evolution with the base bleed intensity in supersonic regime. Results for straight (upper) and rounded (lower) trailing edge geometries.}
    \label{fig:pb-sup}
\end{figure}

\begin{figure}
    \centering
    \includegraphics[width=0.95\textwidth]{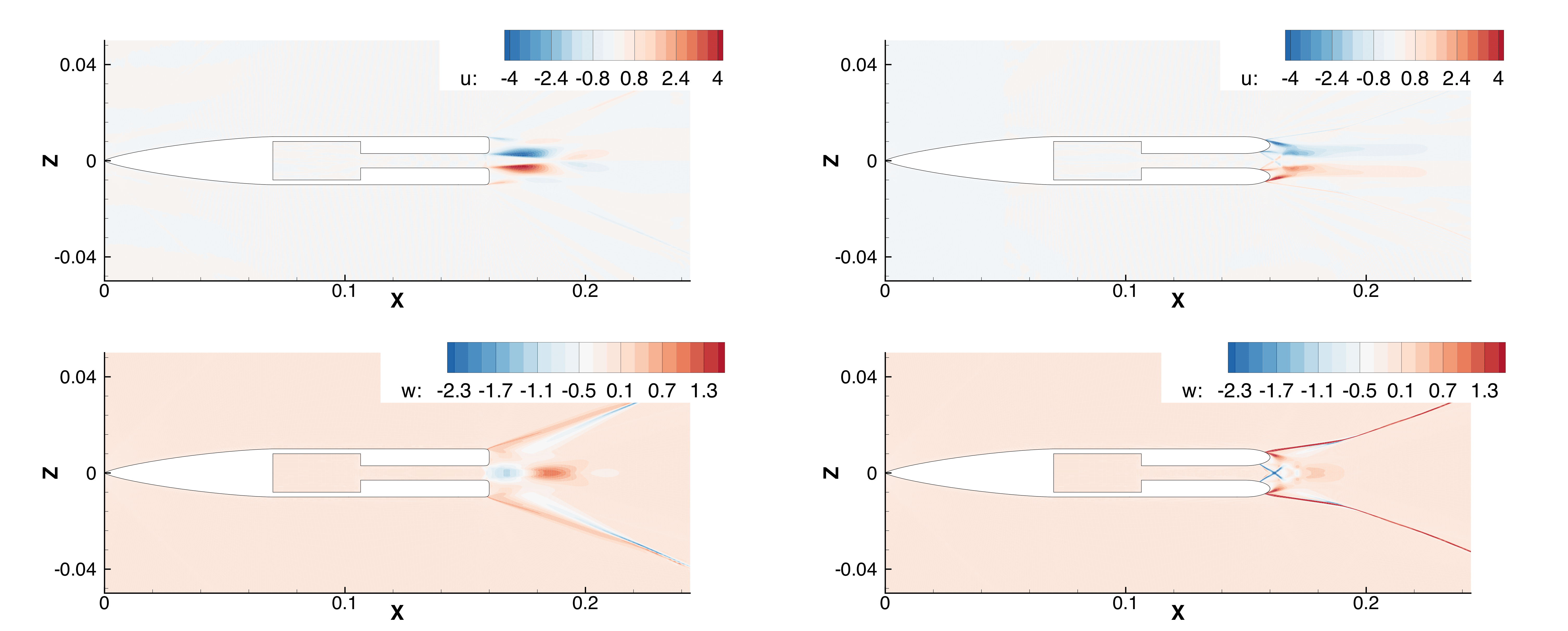}
    \caption{Global mode related with the geometrical sudden expansion, supersonic regime. Streamwise (\textit{u}) and vertical (\textit{w}) velocity perturbations are shown for both trailing edge configurations.}
    \label{fig:sup-globalmode}
\end{figure}

The flow topology, with few sources of unsteadiness, allows the calculation of RANS solutions for the half domain simulations, obtaining steady \textit{base flows} for the stability analysis. The analysis was performed for chosen mass flows, within the blowing rates related with the non-symmetric configuration, revealing again the presence of the expansion mode, in the form of a Pitchfork bifurcation. However, the shape of the global mode is different from what could be observed for the subsonic case, as the perturbations will affect the shear layers and the trailing edge shock waves (Fig. \ref{fig:sup-globalmode}). For this regime, it is also clear the influence of the trailing edge geometry on the shape of the eigenmode, that for a rounded shape it abandons the classic perturbation distribution of sudden expansion geometries of anti-symmetrical lobes to a more stylized and line shaped distribution.

\section{Base region flow sensitivity} \label{sec:sens-study}
Controlling the flow downstream a turbine blade is one of the major challenges nowadays, as secondary flows drive most of the losses in this area. With the information extracted from this numerical analysis, we aim to identify those regions that would be more appropriate for flow control applications, reducing the time on a real design process.

For this we propose two ways of action, looking forward to passive and active flow control. The first is based on the modification or addition of geometry to the existent one in order to mitigate the instability, in this case to force a symmetric base bleed flow regardless the intensity of the jet. The second is however more complex, as we will be looking at the control of the flow at the base region by direct actuation on it. As the non-symmetrical phenomena has been linked to an unstable global mode, the sensitivity analysis of the related eigenvalue will provide valuable information on this topic. 

\begin{figure}
    \centering
    \includegraphics[width=0.95\textwidth]{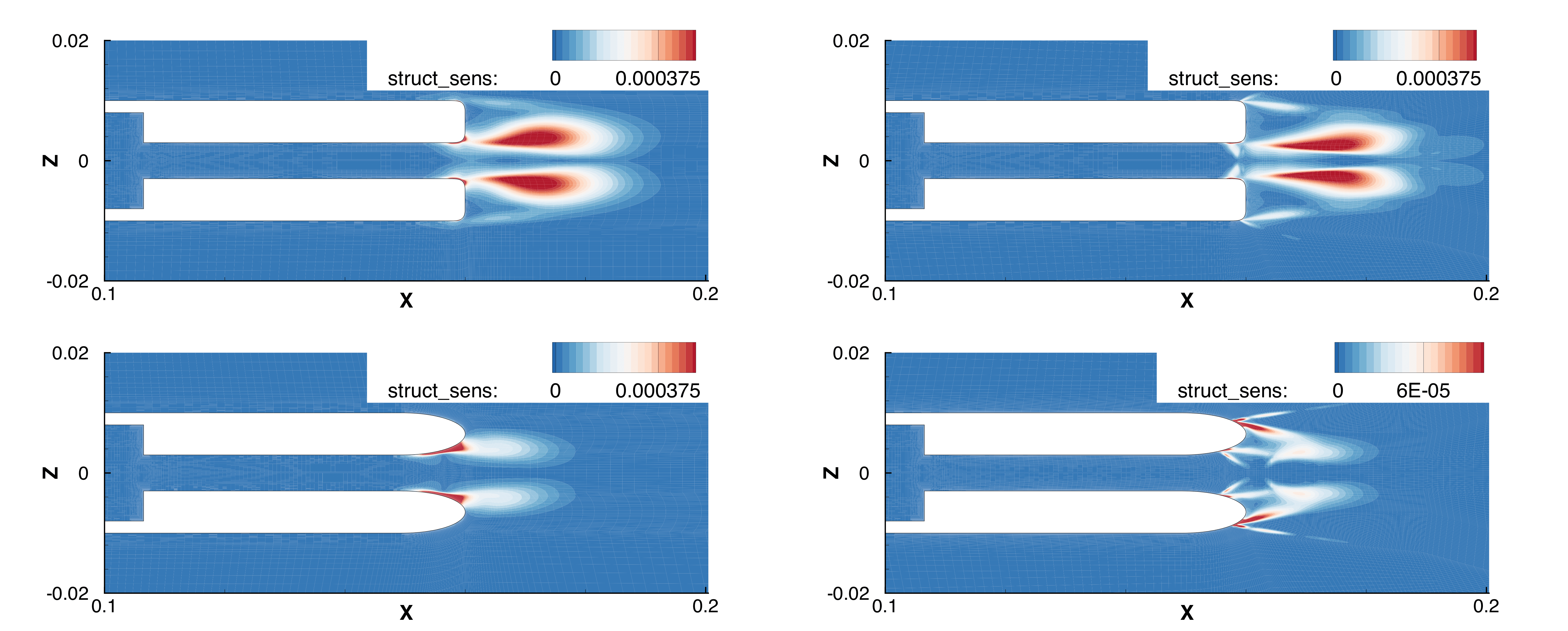}
    \caption{Structural sensitivity of the global mode responsible of the non-symmetrical flow configurations. Left two figures (upper and lower) correspond to subsonic conditions, with the right two figures obtained for supersonic flow.}
    \label{fig:struct-sens}
\end{figure}

The wavemaker regions (core of the instability) are first calculated as the structural sensitivity maps (contours shown on Figure \ref{fig:struct-sens}), showing a strong difference between the straight and the rounded trailing edge shape. For the latest, the core of the instability is always close to the wall, near the flow detachment point in the inner side of the ellipse, with a very different distribution for the supersonic regime where a secondary region appears at the foot of the Prantdtl-Meyer expansion fan. The straight trailing edge presents however the core of the instability downstream the expansion, for both subsonic and supersonic cases, with a secondary sensitive area at the end of the injection channel wall. The structural sensitivity follows the shape of the global modes, being the main core of the instability mainly located at the base region, outside the injection channel. 

As the global modes were obtained for a forced-symmetrical configuration, the sensitivity fields appear symmetric respect to the mid-plane, and they are effective regardless the direction of deflection of the flow. For all the cases, the core of the instability is contained within the recirculation area of the base region. This information is valuable not only to understand where are located the principal mechanisms of the instability, but to consider future stability analysis where a more expensive computational domain is required. Giannetti and Luchini\cite{Giannetti2007} and Sanvido et al.\cite{Sanvido2017} showed that to ensure a good accuracy on the calculation of the eigenvalues, the computational domain for the stability analysis should confine the complete instability core.

\begin{figure}
    \centering
    \includegraphics[width=0.95\textwidth]{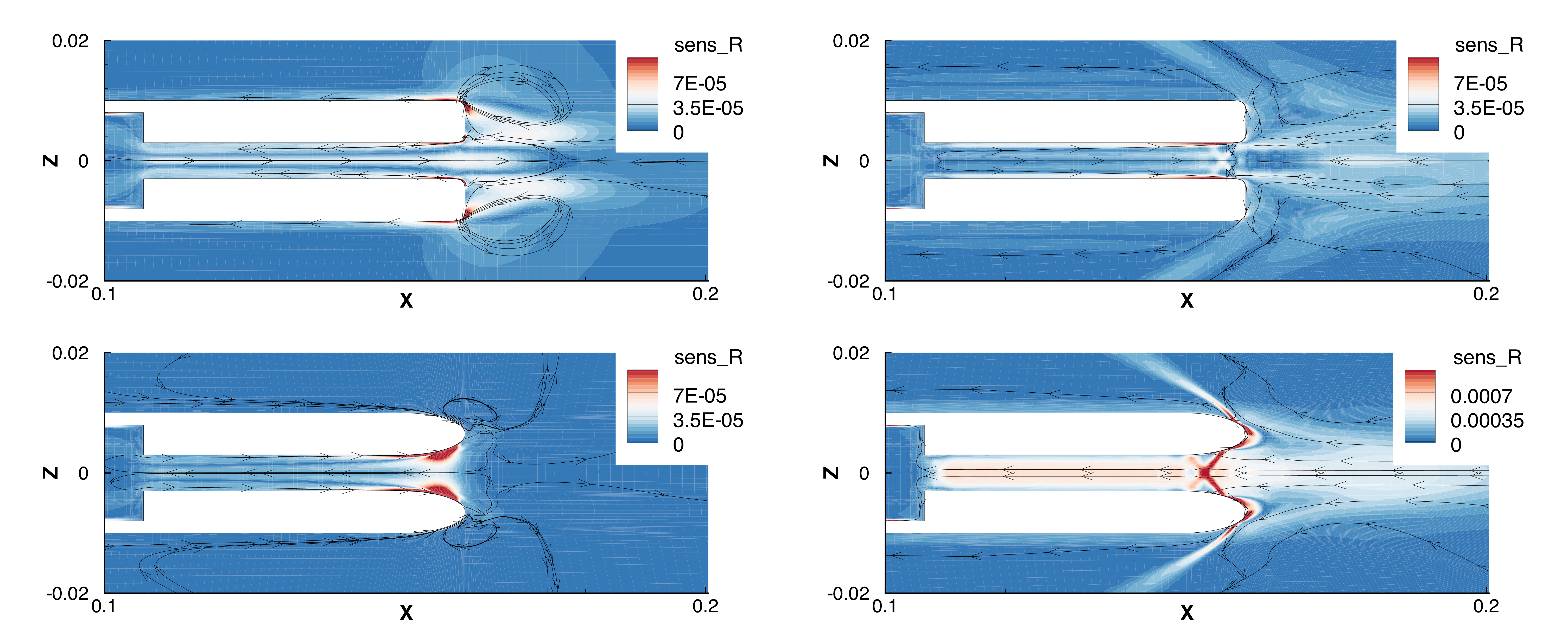}
    \caption{Sensitivity to the application of a steady force. Perturbations in the direction of the streamlines would have a destabilizing effect on the eigenvalue. Left two figures (upper and lower) correspond to subsonic conditions, with the right two figures obtained for supersonic flow.}
    \label{fig:forcing-sens}
\end{figure}

One of the most interesting results that can be extracted from the sensitivity analysis, are the vector field maps related to the sensitivity of the global mode (and its related eigenvalue) to the application of a localized steady forcing. Due to the nature of the instability (with no associated non-frequency), this information could be very valuable in terms of active flow control. Both maps for the subsonic and supersonic configurations are plotted on Fig. \ref{fig:forcing-sens}, with the depicted streamlines indicating the direction in which a steady force would have a destabilizing effect on the global mode. For a straight trailing edge the most sensitive areas remain inside the cooling flow injection channel, very close to the walls and located near the exit. The subsonic regime appears to have a secondary area of sensitivity at the end of the trailing edge, where the main shear layers of the base region are formed. For a rounded trailing edge, however, the most sensitive regions to steady forcing are more elongated, with its core located on the area behind the separation point on the opposite side where the \textit{Coanda} effect is taking place. In supersonic regime, an additional region in the center of the channel appears, related with the expansion of the injected flow.

\begin{figure}
    \centering
    \includegraphics[width=0.95\textwidth]{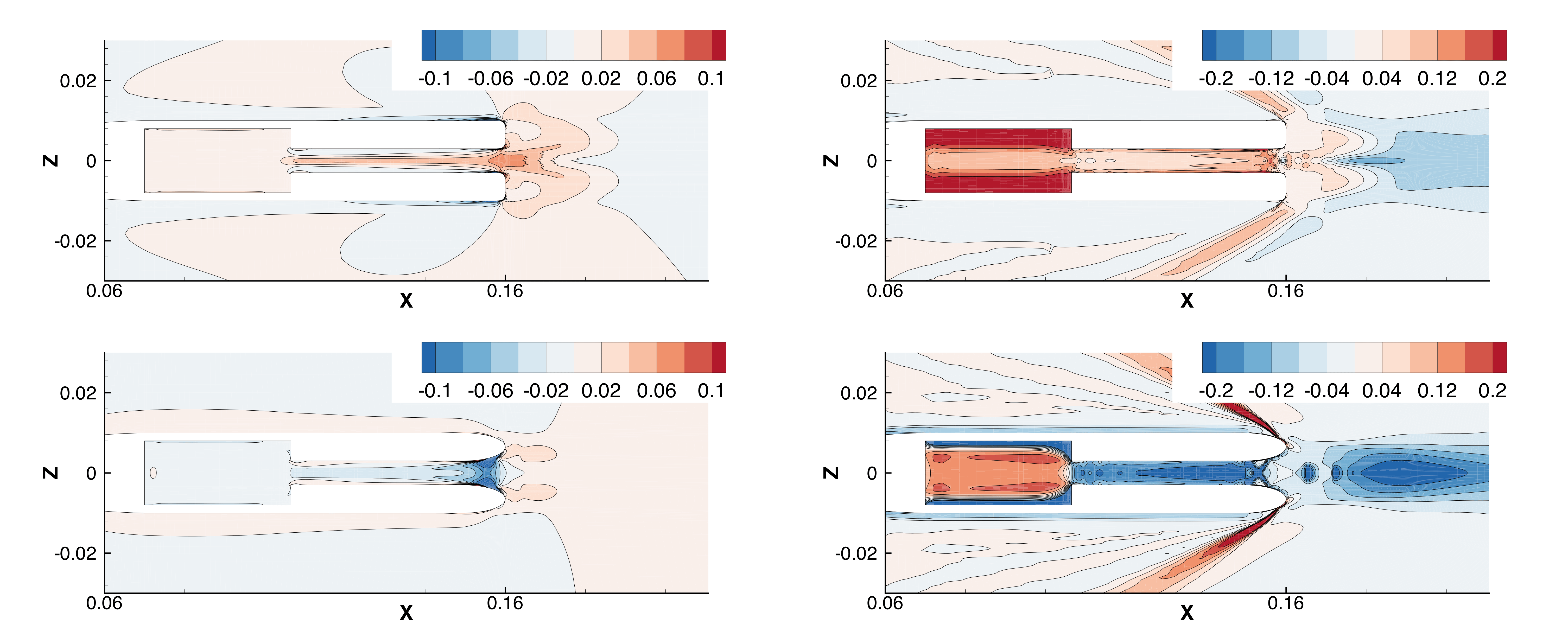}
    \caption{Localization of regions where the location of a small control cylinder will generate a major drift on the eigenvalue. Left two figures (upper and lower) correspond to subsonic conditions, with the right two figures obtained for supersonic flow.}
    \label{fig:passive-sens}
\end{figure}

Following the work of Marquet et al.\cite{Marquet2008}, it is possible to calculate the sensitivity of the eigenvalue to the addition of a small control body to the flow field to control the instability. This is done by assuming a steady force in the direction of the flow, namely the drag force of a small cylinder with diameter \textit{$d^*$}. If this analysis is performed near the bifurcation point, without loose of generality one can say that the sensitivity of the eigenvalue would be:
\begin{equation}
    \nabla \sigma_p = - \alpha ||\mathbf{\bar{q}}(x_0,y_0)|| \nabla_{\mathbf{q_f}} \sigma \cdot \mathbf{\bar{q}}(x_0,y_0)
\end{equation}
where coefficient $\alpha$, constrained as $0 < \alpha \ll 1$, that normally takes the form $\alpha = \frac{1}{2}d^* C_D (Re_{d^*})$.
If this analysis is conducted for the global modes responsible of the change of direction of the flow, the results are similar to those obtained by Fani et al.\cite{Fani2012}, where it was shown that a small cylinder placed at the ``sweet spot" at the end of the channel before a geometrical sudden expansion can mitigate or neutralize the apparition of the instability. Sensitivity maps are shown on Fig. \ref{fig:passive-sens}, where both flow regimes and trailing edge shapes are represented. Interestingly, the rounded trailing edge seems to be far more sensitive to passive control than its reciprocal configuration. For subsonic conditions, the presence of an additional body (or modification of the surface to generate additional drag) at the end of the base bleed channel will have a stabilizing effect on the eigenvalue. Inside a supersonic flow, this area extends upstream the channel, with two additional lobes downstream where a control cylinder could be placed. The analysis for passive control on the straight configurations, show that only specific locations will be suitable for the suppression of the instability by adding a control body.

\section{Conclusions and way forward} \label{sec:conclusions}
The non-symmetrical flow configurations shown here can appear at the base region of a base bleed trailing edge, and are related to a fundamental phenomenon that is normally ignored on the big picture of turbine blade design. This phenomenon has an effect on the development and behaviour of secondary flows, and with the correct information could be exploited for better performance, or cancelled to avoid any undesired interferences.

With the aim of extending the available information in the literature regarding base region flow modulation by trailing edge blowing using coolant flow, a range of subsonic, transonic and supersonic cases have been analyzed. A combination of unsteady RANS simulations and stability analysis has been used to identify those blowing rate intensities for which base pressure presents a maximum value, and also which of these values are related with the apparition of a mild or full \textit{Coanda} effect at the cooling injection channel. The global stability analysis revealed the presence of a global mode connected to the geometrical sudden expansion, that eventually drives and control the direction of the base bleed jet flow. It has been shown that a rounded trailing edge tip provides a less abrupt behavior on the base region pressure changes, accompanied by a higher flow deflection when the first instability of the sudden expansion geometry becomes unstable. The use of rounded edges, moreover, seemed to neutralize the decay in base pressure that was observed for subsonic and transonic flows, as well as the related high frequency vortices that were identified for subsonic and supersonic regimes for the studied configurations.

The information extracted on the study, combined with the results of the sensitivity analysis based on the unstable global modes, will open the door to active and passive flow control techniques to modulate or neutralize the global instability present at the cooling channel, that is directly related with the apparition of a \textit{Coanda} effect at this region. The addition of a small control body at the calculated ``sweet spot" will keep the flow symmetric for a large range of base bleed flow intensities.



\section*{Acknowledgments}
This research has being carried out under the project SSeMID (\textit{Stability and Sensitivity Methods for Industrial Design}, http://ssemid-itn.eu/), which has received funding from the European Union Horizon2020 research and innovation programme under the Marie Skłodowska-Curie grant agreement No 675008. 

The authors acknowledge the computer resources and technical assistance provided by the Centro de Supercomputación y Visualización de Madrid (CeSViMa).

The authors would also like to express their gratitude to the Purdue Experimental Turbine Aero-Thermal Lab of Purdue University for their support during the elaboration of this research.


\bibliography{mendeley}

\end{document}